\documentclass[pra,aps,twocolumn,twoside]{revtex4}
\usepackage{amssymb,amsthm,amsmath}
\theoremstyle{plain}
\newtheorem{theorem}{Theorem}
\newtheorem{lemma}[theorem]{Lemma}
\newtheorem{corollary}[theorem]{Corollary}
\newtheorem{proposition}[theorem]{Proposition}
\newtheorem{remark}[theorem]{Remark}
\newtheorem{conjecture}[theorem]{Conjecture}
\theoremstyle{definition}
\newtheorem{definition}[theorem]{Definition}
\newtheorem{example}[theorem]{Example}
\newcommand{\ket}[1]{|#1 \rangle}
\newcommand{\bra}[1]{\langle #1|}
\newcommand{\proj}[1]{\ket{#1}\!\bra{#1}}
\newcommand{\be}{\begin{equation}}
\newcommand{\ee}{\end{equation}}
\newcommand{\bea}{\begin{eqnarray*}}
\newcommand{\eea}{\end{eqnarray*}}
\newcommand{\h}[1]{\mathcal{H}_{#1} }
\newcommand{\Tr}{{\operatorname{Tr}\,}}
\newcommand{\id}{{\openone}}

\newcommand{\idmap}{{\operatorname{id}}}
\def\opone{\leavevmode\hbox{\small1\kern-3.8pt\normalsize1}}
\newcommand{\E}{{E_{sq}}}

\begin{document}

\title{``Squashed Entanglement'' -- An Additive Entanglement Measure}
\author{Matthias Christandl}
\email{matthias.christandl@qubit.org} \affiliation{Centre for Quantum
Computation, Department of Applied Mathematics and Theoretical Physics,
University of Cambridge, Wilberforce Road, Cambridge CB3 0WA, United Kingdom}
\author{Andreas Winter}
\email{a.j.winter@bris.ac.uk} \affiliation{School of Mathematics, University
of Bristol, University Walk,  Bristol BS8 1TW, United Kingdom}
\date{18th November 2003}

\begin{abstract}
 In this paper, we present a new entanglement monotone for bipartite quantum states. Its
 definition is
 inspired by the so--called intrinsic information of classical cryptography and
 is given by the halved minimum quantum conditional mutual information over all tripartite
 state extensions. We derive certain properties of the new measure which we
 call ``squashed entanglement'': it is a lower bound on
 entanglement of formation and an upper bound on distillable
 entanglement. Furthermore, it is convex, additive on tensor products, and
 superadditive in general.
 \par
 Continuity in the state is the only property of our entanglement measure which
 we cannot provide a proof for.
 We present some evidence, however, that our quantity has this property,
 the strongest indication being a conjectured Fannes type inequality
 for the conditional von Neumann entropy. This inequality is proved in the classical case.
\end{abstract}

\maketitle

\section{Introduction}
\label{sec:intro}
Ever since Bennett et al. \cite{BBPSSW, BDSW} introduced the entanglement
measures of \emph{distillable entanglement} and \emph{entanglement of
formation} in order to measure the amount of nonclassical correlation in a
bipartite quantum state, there has been an interest in an axiomatic approach to
entanglement measures. One natural axiom is LOCC--monotonicity, which means
that an entanglement measure should not increase under \emph{Local Operations
and Classical Communication}. Furthermore, every entanglement measure should
vanish on the set of separable quantum states; it should be convex, additive,
and a continuous function in the state. Though several entanglement measures
have been proposed, it turns out to be difficult to find measures that satisfy
all of the above axioms. One unresolved question is whether or not entanglement
of formation is additive. This is an important question and has recently been
connected to many other additivity problems in quantum information
theory~\cite{Shor}. Other examples are distillable entanglement, which shows
evidence of being neither additive nor convex \cite{ShorSmolinTerhal}, and
\emph{relative entropy of entanglement} \cite{vedral:ent}, which can be proved
to be nonadditive \cite{VollbrechtWerner}.

\par
In this paper we present a functional called ``squashed entanglement'' which
has many of these desirable properties: it is convex, additive on tensor
products and superadditive in general. It is upper bounded by entanglement
cost, lower bounded by distillable entanglement, and we are able to present
some evidence of continuity.

The remaining sections are organised as follows: in section~\ref{sec:squashed}
we will define squashed entanglement and prove its most important properties.
In section~\ref{sec:intrinsic} we will explain its analogy to a quantity called
\emph{intrinsic information}, known from classical cryptography. This
constitutes the motivation for our definition.

The only property that we could find proof for is continuity. A detailed
discussion of this problem follows in section~\ref{sec:continuity}, where we
show that squashed entanglement is continuous on the interior of the set of
states and where we provide evidence in favour of continuity in general. This
evidence is based on a Fannes type inequality for the conditional von Neumann
entropy. It is conjectured in general and is true in the classical case,
which we will prove in the appendix.

\section{Squashed entanglement}
\label{sec:squashed}
In this paper all Hilbert spaces are assumed to be finite dimensional, even
though some of the following definitions and statements make sense also in
infinite dimension.
\par
\begin{definition}
  \label{defi:sq}
  Let $\rho^{AB}$ be a quantum state on a bipartite Hilbert space
  $\h{}=\h{A}\otimes \h{B}$. We define the \emph{squashed entanglement} of
  $\rho^{AB}$ by
  $$\E(\rho^{AB}):=\inf\left\{ \frac{1}{2} I(A;B|E) :
                         \rho^{ABE} \textrm{ extension of } \rho^{AB} \right\} \!.$$
  The infimum is taken over all extensions of $\rho^{AB}$, i.e.~over all
  quantum states $\rho^{ABE}$ with $\rho^{AB}=\Tr_E\rho^{ABE}$.
  $I(A;B|E):=S(AE)+S(BE)-S(ABE)-S(E)$ is the quantum conditional mutual
  information of $\rho^{ABE}$~\cite{CerfAdami}. $\rho^A$ stands for the
  restriction of the state $\rho^{ABE}$ to subsystem $A$, and $S(A)=S(\rho^A)$ is the von Neumann entropy
  of the underlying state, if it is clear from the context. If not, we emphasise the state in subscript,
  $S(A)_\rho$. Note that the dimension of $E$ is \emph{a priori} unbounded.
\end{definition}
\par
Tucci~\cite{Tucci} has previously defined the same functional (without the
factor $\frac{1}{2}$) in connection with his investigations into the
relationship between quantum conditional mutual information and entanglement
measures, in particular entanglement of formation.
\par
Our name for this functional comes from the idea that the right choice of a
conditioning system reduces the quantum mutual information between $A$ and
$B$, thus ``squashing out'' the non--quantum correlations. See
section~\ref{sec:intrinsic} for a similar idea in classical cryptography, which
motivated the above definition.
\begin{example}
  \label{expl:pure}
  Let $\rho^{AB}=\proj{\psi}^{AB}$ be a pure state.
  All extensions of $\rho^{AB}$ are of the form $\rho^{ABE}=\rho^{AB}\otimes\rho^E$;
  therefore
  $$\frac{1}{2}I(A;B|E) = S(\rho^A) = E(\ket{\psi}),$$
  which implies $\E(\proj{\psi}) = E(\ket{\psi})$.
\end{example}
\par
\begin{proposition}
  \label{monotone}
  $\E$ is an \emph{entanglement monotone}, i.e.~it does not increase under
  local quantum operations and classical communication (LOCC) and it is convex.
\end{proposition}
\begin{proof}
  According to~\cite{Vidal:2000} it suffices to verify that
  $\E$ satisfies the following two criteria:
  \begin{enumerate}
    \item For any quantum state $\rho^{AB}$ and any unilocal quantum instrument $({\cal E}_k)$
        --- the ${\cal E}_k$ are completely positive maps and their sum is
        trace preserving~\cite{DaviesLewis} --- performed on either subsystem,
        \[ \E(\rho^{AB}) \geq \sum_k p_k \E(\tilde{\rho}^{AB}_k), \]
        where
        \[ p_k                =\Tr {\cal E}_k(\rho^{AB}) \textrm{ and }
           \tilde{\rho}^{AB}_k=\frac{1}{p_k}{\cal E}_k(\rho^{AB}). \]
    \item $\E$ is convex, i.e.~for all $0\leq \lambda\leq 1$,
        \begin{equation*}\begin{split}
          \E&\left( \lambda \rho^{AB}+(1-\lambda)\sigma^{AB} \right)   \\
            &\phantom{====}
                      \leq  \lambda \E(\rho^{AB})+(1-\lambda)\E(\sigma^{AB}).
        \end{split}\end{equation*}
  \end{enumerate}
  In order to prove 1, we modify the proof of theorem 11.15 in \cite{NielsenChuang} for
  our purpose. By symmetry we may assume that the instrument $({\cal E}_k)$
  acts unilocally on $A$. Now, attach two ancilla systems $A'$ and $A''$
  in states $\ket{0}^{A'}$ and $\ket{0}^{A''}$ to the
  system $ABE$ (i). To implement the quantum operation
  \[ \rho^{ABE} \rightarrow \tilde{\rho}^{AA'BE}:=\sum_k ({\cal E}_k\otimes\idmap_E)(\rho^{ABE})
                                                                          \otimes \proj{k}^{A'}, \]
  with $(\ket{k}^{A'})_k$ being an orthonormal basis on $A'$,
  we perform (ii) a unitary transformation $U$ on $AA'A''$ followed by (iii) tracing
  out the system $A''$. Here, $\tilde{i}$ denotes the system $i \in \{A, B, AB\}$
  after the unitary evolution $U$. Then, for any extension of $\rho^{AB}$,
  \begin{align*}
    I(A;B|E) &\stackrel{\text{(i)}}{=}      I(AA'A'';B|E)                                         \\
             &\stackrel{\text{(ii)}}{=}     I(\tilde{A}\tilde{A'}\tilde{A''};\tilde{B}|\tilde{E}) \\
             &\stackrel{\text{(iii)}}{\geq} I(\tilde{A}\tilde{A'};\tilde{B}|\tilde{E})            \\
             &\stackrel{\text{(iv)}}{=}     I(\tilde{A'};\tilde{B}|\tilde{E})
                                              +I(\tilde{A};\tilde{B}|\tilde{E}\tilde{A'})         \\
             &\stackrel{\text{(v)}}{\geq}   \sum_k p_k I(\tilde{A};\tilde{B}|\tilde{E})_{\rho_k}  \\
             &\stackrel{\text{(vi)}}{\geq}  \sum_k 2 p_k \E(\rho_k).
  \end{align*}
  The justification of these steps is as follows:
  attaching auxiliary pure systems does not change the entropy of a system,
  step (i). The unitary evolution affects only the systems $AA'A''$ and
  therefore does not affect the quantum conditional mutual information in step (ii). To
  show that discarding quantum systems cannot increase the quantum conditional mutual information
  \[ I(\tilde{A}\tilde{A'};\tilde{B}|\tilde{E}) \leq
              I(\tilde{A}\tilde{A'}\tilde{A''};\tilde{B}|\tilde{E}) \]
  we expand it into
  \begin{equation*}\begin{split}
    &S(AA'E)+S(BE)-S(AA'BE)-S(E)                  \\
    &\phantom{==}
     \leq S(AA'A''E)+S(BE)-S(AA'A''BE)-S(E),
  \end{split}\end{equation*}
  which is equivalent to
  \[ S(AA'E)-S(AA'BE) \leq S(AA'A''E)-S(AA'A''BE), \]
  the strong subadditivity~\cite{LiebRuskai};
  this shows step (iii), and for step (iv) we use the \emph{chain rule},
  $$I(XY;Z|U) = I(X;Z|U)+I(Y;Z|UY).$$
  For step (v), note that the first term, $I(\tilde{A'};\tilde{B}|\tilde{E})$, is non--negative
  and that the second term, $I(\tilde{A};\tilde{B}|\tilde{E}\tilde{A'})$, is
  identical to the expression in the next line.
  Finally, we have (vi) since $\rho_k^{\tilde{A}\tilde{B}\tilde{E}}$ is a valid
  extension of $\rho_k$. As the original extension of $\rho^{AB}$ was arbitrary, the
  claim follows.
  \par
  To prove convexity, property 2, consider any extensions $\rho^{ABE}$
  and $\sigma^{ABE}$ of the states
  $\rho^{AB}$ and $\sigma^{AB}$, respectively. It is clear that we can assume,
  without loss of generality,
  that the extensions are defined on identical systems $E$.
  Combined, $\rho^{ABE}$ and $\sigma^{ABE}$ form an extension
  $$\tau^{ABEE'} := \lambda\rho^{ABE}\otimes\proj{0}^{E'}
                      + (1-\lambda)\sigma^{ABE}\otimes\proj{1}^{E'}$$
  of the state $\tau^{AB}=\lambda\rho^{AB}+(1-\lambda)\sigma^{AB}$. The convexity of
  squashed entanglement then follows from the observation
  \begin{equation*}\begin{split}
    \lambda I(A;B|E)_{\rho} + &(1-\lambda) I(A;B|E)_{\sigma}\\
                              &=    I(A;B|EE')_{\tau} \geq 2\E(\tau^{AB}).
  \end{split}\end{equation*}
\end{proof}
\par
\begin{proposition}
  \label{additive}
  $\E$ is superadditive in general, and additive on tensor products, i.e.
  \[ \E(\rho^{AA'BB'}) \geq \E(\rho^{AB}) + \E(\rho^{A'B'}) \]
  is true for every density operator $\rho^{AA'BB'}$ on
  $\h{A} \otimes \h{A'} \otimes \h{B} \otimes \h{B'}$,
  $\rho^{AB}=\Tr_{A'B'} \rho^{AA'BB'}$, and
  $\rho^{A'B'}=\Tr_{AB} \rho^{AA'BB'}$.
  \[ \E(\rho^{AA'BB'}) = \E(\rho^{AB}) + \E(\rho^{A'B'}) \]
  for $\rho^{AA'BB'}=\rho^{AB} \otimes \rho^{A'B'}$.
\end{proposition}
\begin{proof}
  We start with superadditivity and assume that
  $\rho^{AA'BB'E}$ on $\h{A} \otimes \h{A'} \otimes \h{B} \otimes \h{B'} \otimes
  \h{E}$ is an extension of $\rho^{AA'BB'}$, i.e. $\rho^{AA'BB'}= \Tr_E \rho^{AA'BB'E}$.
  Then,
  \begin{align*}
     I(AA';BB'|E)&=    I(A;BB'|E)+I(A';BB'|EA)  \\
                 &=    I(A;B|E)+I(A;B'|EB)      \\
                 &\phantom{=}
                      +I(A';B'|EA)+I(A';B|EAB') \\
                 &\geq I(A;B|E)+I(A';B'|EA)     \\
                 &\geq 2\E(\rho^{AB})+2\E(\rho^{A'B'})
  \end{align*}
  The first inequality is due to strong subadditivity of the von Neumann
  entropy. Note that $E$ is an extension for system $AB$ and $EA$ extends system
  $A'B'$. Hence, the last inequality is true since squashed entanglement
  is defined via the infimum over all extensions of the respective states.
  The calculation is independent of the choice of the extension, which proves
  superadditivity.
  \par
  A special case of the above is superadditivity on product states
  $\rho^{AA'BB'}:=\rho^{AB} \otimes \rho^{A'B'}$. To conclude that
  $\E$ is indeed additive on tensor products, it therefore suffices
  to prove subadditivity on tensor products.
  \par
  Let $\rho^{ABE}$ on $\h{A} \otimes \h{B} \otimes \h{E}$ be an extension of
  $\rho^{AB}$ and let $\rho^{A'B'E'}$ on  $\h{A'} \otimes \h{B'} \otimes
  \h{E'}$ be an extension for $\rho^{A'B'}$. It is evident that
  $\rho^{ABE} \otimes \rho^{A'B'E'}$ is a valid extension for
  $\rho^{AA'BB'} = \rho^{AB} \otimes \rho^{A'B'}$, hence
  \begin{align*}
    2\E(\rho^{AA'BB'}) &\leq I(AA';BB'|EE')                                 \\
                       &=    I(A;B|EE')+\underbrace{I(A;B'|EE'B)}_{=0}      \\
                       &\phantom{=}
                            +I(A';B'|EE'A)+\underbrace{I(A';B|EE'AB')}_{=0} \\
                       &=    I(A;B|E)+I(A';B'|EE'A)                         \\
                       &=    I(A;B|E)+I(A';B'|E').
  \end{align*}
  This inequality holds for arbitrary extensions of $\rho^{AB}$ and $\rho^{A'B'}$.
  We therefore conclude that $\E$ is subadditive on tensor products.
\end{proof}
\par
\begin{proposition}
  \label{EoF}
  $\E$ is upper bounded by \emph{entanglement of formation~\cite{BBPSSW,BDSW}}:
  $$\E(\rho^{AB}) \leq E_F(\rho^{AB}).$$
\end{proposition}
\begin{proof}
  Let $\{p_k, \ket{\Psi_k}\}$ be a pure state ensemble for $\rho^{AB}$:
  $$\sum_k p_k \proj{\Psi_k}^{AB} = \rho^{AB}.$$
  The purity of the ensemble implies
  \[ \sum_k p_k S(A)_{\Psi_k} = \frac{1}{2} \sum_k p_k I(A;B)_{\Psi_k}. \]
  Consider the following extension $\rho^{ABE}$ of $\rho^{AB}$:
  \[ \rho^{ABE}:= \sum_k p_k \proj{\Psi_k}^{AB} \otimes \proj{k}^E. \]
  It is elementary to compute
  \[\sum_k p_k S(A)_{\Psi_k} = \frac{1}{2} \sum_k p_k I(A;B)_{\Psi_k}
                             = \frac{1}{2} I(A;B|E). \]
  Thus, it is clear that entanglement of formation can be regarded as an
  infimum over a certain class of extensions of $\rho^{AB}$. Squashed
  entanglement is an infimum over \emph{all} extensions of $\rho^{AB}$,
  evaluated on the same quantity $\frac{1}{2}I(A;B|E)$ and therefore smaller or
  equal to entanglement of formation.
\end{proof}
\par
\begin{corollary}
  \label{E-cost}
  $\E$ is upper bounded by \emph{entanglement cost}:
  $$\E(\rho^{AB}) \leq E_C(\rho^{AB}).$$
\end{corollary}
\begin{proof}
  Entanglement cost is equal to the regularised entanglement of
  formation~\cite{HaydenHorodeckiTerhal:2000},
  \[ E_C(\rho^{AB})=\lim_{n\rightarrow\infty} \frac{1}{n}E_F\left( (\rho^{AB})^{\otimes n}\right). \]
  This, together with proposition \ref{EoF}, and the additivity of
  the squashed entanglement (proposition \ref{additive}) implies
  \begin{align*}
    E_C(\rho^{AB}) &=    \lim_{n\rightarrow\infty}
                                 \frac{1}{n}E_F\left( (\rho^{AB})^{\otimes n}\right) \\
                   &\geq \lim_{n\rightarrow\infty}
                                 \frac{1}{n}\E\left( (\rho^{AB})^{\otimes n}\right)  \\
                   &=    \E(\rho^{AB}).
  \end{align*}
\end{proof}
\par

\begin{theorem}
  \label{value-zero}
  Squashed entanglement vanishes for every separable density matrix
  $\rho^{AB}$, i.e.
  \[ \rho^{AB} \textrm{separable} \implies \E(\rho^{AB})=0.\]
  Conversely, if there exists a finite extension $\rho^{ABE}$ of
  $\rho^{AB}$ with vanishing quantum conditional mutual information, then
  $\rho^{AB}$ is separable, i.e.
  \[ I(A;B|E)=0 \textrm{ and } \dim \h{E} <\infty \implies
  \rho^{AB} \textrm{ separable}.\]
\end{theorem}
\begin{proof}
Every separable $\rho^{AB}$ can be written as a convex combination of
separable pure states,
\[  \rho^{AB}=\sum_i p_i \proj{\psi_i}^A \otimes
    \proj{\phi_i}^B.
\]
The quantum mutual conditional information of the extension
\[ \rho^{ABE}:=\sum_i p_i \proj{\psi_i}^A \otimes
    \proj{\phi_i}^B \otimes \proj{i}^E \],
with orthonormal states $(\ket{i}^E)$, is zero. Squashed entanglement thus
vanishes on the set of separable states.

To proof the second assertion assume that there exists an extension
$\rho^{ABE}$ of $\rho^{AB}$ with $I(A;B|E)=0$ and $\dim \h{E} < \infty$.
Now, a recently obtained result~\cite{SSA-eq} on the structure of such states
$\rho^{ABE}$ applies, and as a corollary $\rho^{AB}$ is separable.
\end{proof}

\begin{remark}
    The minimisation in squashed entanglement ranges over extensions of
    $\rho^{AB}$ with \emph{a priori} unbounded size. $\E(\rho)=0$ is thus possible,
    even if any finite extension has strictly positive quantum conditional mutual information.
    Therefore, without a bound on the dimension of the extending system,
    the second part of theorem \ref{value-zero} does not suffice to conclude that
    $\E(\rho^{AB})$ implies separability of $\rho^{AB}$.
    A different approach to this question could be provided by a possible approximate version
    of the main result of~\cite{SSA-eq}: if there is an extension $\rho^{ABE}$ with
    small quantum conditional mutual information, then the
    $\rho^{AB}$ is close to a separable state.
    For further discussion on this question, see sections~\ref{sec:intrinsic} and~\ref{sec:continuity}.

    Note that the strict positivity of squashed entanglement for entangled states
    would, via corollary~\ref{E-cost}, imply strict positivity of entanglement cost for all entangled
    states. This is not yet proven, but conjectured as a consequence of
    the additivity conjecture of entanglement of formation.
\end{remark}
\par
\begin{example}
  It is worth noting that in general $\E$ is strictly smaller than $E_F$ and $E_C$:
  consider the totally antisymmetric state $\sigma^{AB}$ in a two--qutrit system
  $$\sigma^{AB} = \frac{1}{3}\bigl( \proj{I}+\proj{II}+\proj{III} \bigr),$$
  with
  \begin{align*}
    \ket{I}   &= \frac{1}{\sqrt{2}}\left( \ket{2}^A\ket{3}^B-\ket{3}^A\ket{2}^B \right), \\
    \ket{II}  &= \frac{1}{\sqrt{2}}\left( \ket{3}^A\ket{1}^B-\ket{1}^A\ket{3}^B \right), \\
    \ket{III} &= \frac{1}{\sqrt{2}}\left( \ket{1}^A\ket{2}^B-\ket{2}^A\ket{1}^B \right). \\
  \end{align*}
  On the one hand, it is known from~\cite{yura} that $E_F(\sigma^{AB})=E_C(\sigma^{AB})=1$,
  though, on the other hand, we may consider the trivial extension,
  $$\E(\sigma^{AB}) \leq \frac{1}{2} I(A;B) = \frac{1}{2}\log 3 \approx 0.792.$$
  The best known upper bound on $E_D$ for this state, the
  \emph{Rains bound}~\cite{Rains:PPT},
  gives the only slightly smaller value $\log \frac{5}{3} \approx 0.737$. It remains open if there exist states
  for which squashed entanglement is smaller than the Rains bound.
\end{example}
\par
\begin{proposition}
\label{distillable-bound}
  $\E$ is lower bounded by \emph{distillable entanglement~\cite{BBPSSW,BDSW}}:
  $$E_D(\rho^{AB}) \leq \E(\rho^{AB}).$$
\end{proposition}
\begin{proof}
  Consider any entanglement distillation protocol by LOCC, taking
  $n$ copies of the state $(\rho^{AB})^{\otimes n}$ to a state $\sigma^{AB}$ such that
  \begin{equation}
    \label{norm-difference}
    \bigl\| \sigma^{AB}-\proj{s}^{AB} \bigr\|_1 \leq \delta,
  \end{equation}
  with $\ket{s}$ being a maximally entangled state of Schmidt rank $s$.
  We may assume without loss of generality that the support of $\sigma^A$ and $\sigma^B$
  is contained in the $s$-dimensional support of $\Tr_B\proj{s}$ and
  $\Tr_A\proj{s}$, respectively. Using propositions~\ref{additive} and~\ref{monotone}, we have
  \begin{equation}
    \label{eq:monotone-step}
    n \E(\rho^{AB}) =    \E\left( (\rho^{AB})^{\otimes n} \right)
                    \geq \E(\sigma^{AB}),
  \end{equation}
  so that it is only necessary to estimate $\E(\sigma^{AB})$ versus
  $\E(\proj{s}^{AB})=\log s$ (see example~\ref{expl:pure}).
  For this, let $\sigma^{ABE}$ be an arbitrary extension of
  $\sigma^{AB}$ and consider a purification of it, $\ket{\Psi} \in \h{ABEE'}$.
  Chain rule and monotonicity of the quantum mutual
  information allow us to estimate
  \begin{align*}
    I(A;B|E) &=    I(AE;B) - I(E;B)    \\
             &\geq I(A;B)  - I(EE';AB) \\
             &=    I(A;B)  - 2S(AB).
  \end{align*}
  Further applications of Fannes inequality~\cite{Fannes},
  lemma~\ref{fannes}, give $I(A;B) \geq 2\log s - f(\delta)\log s$
  and $2S(AB) \leq f(\delta)\log s$, with a function $f$ of $\delta$
  vanishing as $\delta$ approaches $0$. Hence
  $$\frac{1}{2} I(A;B|E) \geq \log s - f(\delta)\log s.$$
  Since this is true for all extensions, we can put this together
  with eq.~(\ref{eq:monotone-step}), and obtain
  $$\E(\rho^{AB}) \geq \frac{1}{n} \bigl( 1-f(\delta) \bigr) \log s,$$
  which, with $n\rightarrow\infty$ and $\delta\rightarrow 0$,
  concludes the proof, because we considered an arbitrary distillation
  protocol.
\end{proof}
\par
\begin{remark} \label{pure-continuity}
  In the proof of proposition \ref{distillable-bound} we made use of the
  continuity of $\E$ in the vicinity of maximally entangled states.
  Similarly, $\E$ can be shown to
  be continuous in the vicinity of \emph{any} pure state. This, together with proposition \ref{monotone},
  the additivity on tensor products (second part of proposition \ref{additive}), and the
  normalisation on Bell states, suffices to prove corollary~\ref{E-cost} and
  proposition~\ref{distillable-bound}~\cite{Horodecki-Limits}.
\end{remark}
\par
\begin{corollary}
  \[ \frac{1}{2} \big( I(A;B)-S(AB)\big) \leq \E (\rho^{AB}). \]
\end{corollary}
\begin{proof}
  The recently established \emph{hashing inequality}~\cite{DevetakWinter}
  provides a lower bound for the \emph{one--way distillable entanglement}
  $E_D^{\rightarrow} (\rho^{AB})$,
  $$S(B)-S(AB) \leq E_D^{\rightarrow}(\rho^{AB}).$$
  Interchanging the roles of $A$ and $B$, we have
  $$\frac{1}{2} \big(I(A;B)-S(AB)\big)\leq E_D(\rho^{AB})$$
  where we use the fact that one--way distillable entanglement is smaller or
  equal to distillable entanglement. This, together with the bound from
  proposition~\ref{distillable-bound}, implies the assertion.
\end{proof}

\section{Analogy to intrinsic information}
\label{sec:intrinsic}
Intrinsic information is a quantity that serves as a measure for the
correlations between random variables in information--theoretical secret--key
agreement~\cite{MaurerWolf:1999}. The \emph{intrinsic (conditional mutual)
information} between two discrete random variables $X$ and $Y$, given a third
discrete random variable $Z$, is defined as
\begin{equation*}\begin{split}
  I(X;Y\downarrow Z) = \inf&\left\{ I(X;Y|\bar{Z}) : \bar{Z} \text{ with } \right. \\
                           &\phantom{==:}\left.
                            XY \rightarrow Z \rightarrow \bar{Z} \text{ a Markov chain} \right\}\!.
\end{split}\end{equation*}
The infimum extends over all discrete channels $Z$ to $\bar{Z}$ that are
specified by a conditional probability distribution $P_{\bar{Z}|Z}$.
\par
A first idea to utilise intrinsic information for measuring quantum
correlations was mentioned in~\cite{GisinWolf:2000}. This inspired the
proposal of a \emph{quantum analog to intrinsic
information}~\cite{Christandl:diploma} in which the Shannon conditional
mutual information plays a role similar to the quantum conditional mutual
information in squashed entanglement. This proposal possesses certain good
properties demanded of an entanglement measure, and it opened the discussion
that has resulted in the current work.
\par
Before we state some similarities in the properties that the \emph{intrinsic
information} and \emph{squashed entanglement} have in common, we would like to
stress their obvious relation in terms of the definitions. Let
$\ket{\Psi}^{ABC}$ be a purification of $\rho^{AB}$ and let $\rho^{ABE}$ be an
extension of $\rho^{AB}$ with purification $\ket{\Phi}^{ABEE'}$. Remark that
all purifications of $\rho^{AB}$ are equivalent in the sense that there is a
suitable unitary transformation on the purifying system with
\[ \id^{AB} \otimes U: \ket{\Psi}^{ABC} \mapsto \ket{\Phi}^{ABEE'}. \]
Applying a partial trace operation over system $E'$ then results in the
completely positive map
\begin{eqnarray*}
                 \Lambda &:& {\cal B}(\h{C})   \longrightarrow {\cal B}(\h{E}), \\
  \idmap \otimes \Lambda &:& \proj{\Psi}^{ABC} \longmapsto     \rho^{ABE}.
\end{eqnarray*}
Conversely, every state
$\rho^{ABE}$ constructed in this manner is an extension of $\rho^{AB}$.
\par
This shows that the squashed entanglement equals
\begin{equation}\begin{split}
  \label{squash-alt}
  \E(\rho^{AB})=\inf&\left\{ \frac{1}{2}I(A;B|E) : \right. \\
                            &\phantom{==:}\left.
                            \rho^{ABE}=(\idmap\otimes\Lambda) \proj{\Psi}^{ABC} \right\} \!,
\end{split}\end{equation}
where the infimum includes all quantum operations
$\Lambda:{\cal B}(\h{C}) \rightarrow {\cal B}(\h{E})$.
\par
In~\cite{Christandl:Renner:Wolf} it is shown that the minimisation in
$I(X;Y\downarrow Z)$ can be restricted to random variables
$\bar{Z}$ with a domain equal to that of $Z$. This shows that the infimum in
the definition is in effect a minimum and that the intrinsic information is a
continuous function of the distribution $P_{XYZ}$. It is interesting to note
that the technique used there (and, for that matter, also in the proof that
entanglement of formation is achieved as a minimum over pure state ensembles
$\rho^{AB}=\sum_k p_k \proj{\Psi_k}^{AB}$ of size $({\rm rank}\,\rho^{AB})^2$),
does not work for our problem, and so, we do not have an easy proof of the
continuity of squashed entanglement. In the following section this issue will
be discussed in some more detail.
\par
In the cryptographic context in which it appears, intrinsic information serves
as an upper bound for the secret--key rate
$S=S(X;Y||Z)$~\cite{MaurerWolf:1999}.
$S$ is the rate at which two parties, having access to repeated realisations of
$X$ and $Y$, can distill secret correlations about which a third
party, holding realisations of $Z$, is almost ignorant. This distillation
procedure includes all protocols in which the two parties communicate via a
public authenticated classical channel to which the eavesdropper has access but
cannot alter the transmitted messages. Clearly, one can interpret distillable
entanglement as the quantum analog to the secret--key rate. On the one hand,
\emph{secret quantum correlations}, i.e. maximally entangled states of qubits,
are distilled from a number of copies of $\rho^{AB}$. In the classical
cryptographic setting, on the other hand, one aims at distilling \emph{secret
classical correlations}, i.e. secret classical bits, from a number of
realisations of a triple of random variables $X, Y$ and $Z$.
\par
We proved in proposition \ref{distillable-bound} that squashed entanglement
is an upper bound for distillable entanglement. Hence, it provides a bound in
entanglement theory which is analogous to the one in information--theoretic
secure key agreement, where intrinsic information bounds the secret--key rate
from above.
\par
This analogy extends further to the bound on the formation of quantum states
(proposition \ref{EoF} and corollary \ref{E-cost}), where we know of a recently
proven classical counterpart: namely that the intrinsic information is a lower
bound on the formation cost of correlations of a triple of random variables $X,
Y$ and $Z$ from secret correlations~\cite{RennerWolf:New}.

\section{The question of continuity}
\label{sec:continuity}
Intrinsic information, discussed in the previous section, and entanglement of
formation are continuous functions of the probability distribution and state,
respectively. This is so, because in both cases we are able to restrict the
minimisation to a compact domain; in the case of intrinsic information to
bounded range $\bar{Z}$ and in entanglment of formation to bounded size
decompositions, noting that the functions to be minimised are continuous.
\par
Thus, by the same general principle, we could show continuity if we had a
universal bound $d$ on the dimension of $E$ in definition~\ref{defi:sq}, in
the sense that every value of $I(A;B|E)$ obtainable by general extensions can
be reproduced or beaten by an extension with a $d$-dimensional system $E$.
Note that if this were true, then (just as for intrinsic information and
entanglement of formation) the infimum would actually be a minimum: in
remark~\ref{value-zero} we have explained that then $\E(\rho^{AB})=0$ would
imply, using the result of~\cite{SSA-eq}, that $\rho^{AB}$ is separable.
\par
As it is, we cannot yet decide on this question, but we would like to present a
reasonable conjecture, an inequality of the Fannes type~\cite{Fannes} for the
conditional von Neumann entropy, which we can show to imply continuity of $\E$.
Let us first revisit Fannes' inequality in a slightly nonstandard form:
\begin{lemma}
  \label{fannes}
  For density operators $\rho$, $\sigma$ on the same $d$-dimensional Hilbert
  space, with $\| \rho-\sigma \|_1 \leq \epsilon$,
  $$\bigl| S(\rho)-S(\sigma) \bigr| \leq \eta(\epsilon)+\epsilon\log d,$$
  with the universal function
  \begin{equation*}
    \eta(\epsilon) = \begin{cases}
                       -\epsilon\log\epsilon & \epsilon\leq \frac{1}{4}, \\
                       \frac{1}{2}           & \text{otherwise.}
                     \end{cases}
  \end{equation*}
  Observe that $\eta$ is a concave function. \qed
\end{lemma}
Now we can state the conjecture, recalling that for a density operator
$\rho^{AB}$ on a bipartite system ${\cal H}_A\otimes{\cal H}_B$, the
conditional von Neumann entropy~\cite{CerfAdami} is defined as
$$S(A|B) := S(\rho^{AB})-S(\rho^B).$$
\begin{conjecture}
  \label{cond-fannes}
  For density operators $\rho$, $\sigma$ on the bipartite system
  ${\cal H}_A\otimes{\cal H}_B$, with $\| \rho-\sigma \|_1 \leq \epsilon$,
  $$\bigl| S(A|B)_\rho-S(A|B)_\sigma \bigr| \leq \eta(2\epsilon)+3\epsilon\log d_A,$$
  with $d_A = \dim{\cal H}_A$, or some other universal function $f(\epsilon, d_A)$
  vanishing at $\epsilon=0$ on the right hand side.
\end{conjecture}
Note that the essential feature of the conjectured inequality is that it only
makes reference to the dimension of system $A$. If we were to use Fannes
inequality directly with the definition of the conditional von Neumann entropy,
we would pick up additional terms containing the logarithm of
$d_B=\dim{\cal H}_B$. In the appendix we show that this conjecture is true in
the classical case, or more precisely, in the more general case where the
states are classical on system $B$.
\par\medskip
In order to show that the truth of this conjecture implies continuity of
$\E$, consider two states $\rho^{AB}$ and $\sigma^{AB}$ with
$\| \rho^{AB}-\sigma^{AB} \|_1 \leq \epsilon$. By well-known relations
between fidelity and trace distance~\cite{Fuchs:vandeGraaf} this means that
$F\bigl( \rho^{AB},\sigma^{AB} \bigr) \geq 1-\epsilon$,
hence~\cite{Jozsa,Uhlmann} we can find purifications $\ket{\Psi}^{ABC}$ and
$\ket{\Phi}^{ABC}$ of $\rho^{AB}$ and $\sigma^{AB}$, respectively, such that
$F\bigl( \ket{\Psi}^{ABC},\ket{\Phi}^{ABC} \bigr) \geq 1-\epsilon$.
Using~\cite{Fuchs:vandeGraaf} once more, we get
$$\bigl\| \proj{\Psi}^{ABC}-\proj{\Phi}^{ABC} \bigr\|_1 \leq 2\sqrt{\epsilon}.$$
Now, let $\Lambda$ be any quantum operation as in eq.~(\ref{squash-alt}):
it creates extensions of $\rho^{AB}$ and $\sigma^{AB}$,
\begin{align*}
  \rho^{ABE}   &= (\idmap\otimes\Lambda)\proj{\Psi}^{ABC}, \\
  \sigma^{ABE} &= (\idmap\otimes\Lambda)\proj{\Phi}^{ABC},
\end{align*}
with
$$\bigl\| \rho^{ABE}-\sigma^{ABE} \bigr\|_1 \leq 2\sqrt{\epsilon}.$$
Hence, using $I(A;B|E)=S(A|E)+S(B|E)-S(AB|E)$, we can estimate
\begin{equation*}\begin{split}
  \bigl| I(A;B|E)_\rho - I(A;B|E)_\sigma \bigr|
                                      &\leq \bigl| S(A|E)_\rho - S(A|E)_\sigma \bigr|   \\
                                      &\phantom{=}
                                           +\bigl| S(B|E)_\rho - S(B|E)_\sigma \bigr|   \\
                                      &\phantom{=}
                                           +\bigl| S(AB|E)_\rho - S(AB|E)_\sigma \bigr| \\
                                      &\leq 3\eta\bigl(2\sqrt{\epsilon}\bigr)
                                           +6\sqrt{\epsilon}\log(d_A d_B)               \\
                                      &=:\epsilon'.
\end{split}\end{equation*}
Since this applies to any quantum operation $\Lambda$ and thus to every state
extension of $\rho^{AB}$ and $\sigma^{AB}$, respectively, we obtain
$$\bigl| \E(\rho^{AB})-\E(\sigma^{AB}) \bigr| \leq \epsilon',$$
with $\epsilon'$ universally dependent on $\epsilon$ and vanishing
with $\epsilon\rightarrow 0$.
\qed

\begin{remark}
  Since $\E$ is convex it is trivially upper semicontinuous.
  This also follows from the fact that squashed entanglement is
  an infimum of continuous functions obtained by bounding the size of the
  dimension of system $E$.
\end{remark}

This observation, together with results from the general theory of convex
functions, implies that squashed entanglement is continuous ``almost
everywhere''. Specifically, with theorem~10.1 in \cite{rockafeller}, we have:

\begin{proposition}
  \label{prop:facewise:cont}
  $\E$ is continuous on the interior of the set of states (i.e. on the faithful states), and
  more generally, it is continuous when restricted to the relative interior
  of all faces of the state set.
  \par
  Continuity near pure states (see remark \ref{pure-continuity}) thus implies
  continuity of $\E$ on the set of all rank--$2$ density operators.
  \qed
\end{proposition}

\section{Conclusion}
\label{sec:conclusion}
In this paper we have presented a new measure of entanglement, which by its
very definition allows for rather simple proofs of monotonicity under LOCC,
convexity, additivity for tensor products and superadditivity in general, all
by application of the strong subadditivity property of quantum entropy. We
showed the functional, which we call ``squashed entanglement'', to be lower
bounded by the distillable entanglement and upper bounded by the entanglement
cost. Thus, it has most of the ``good'' properties demanded by the axiomatic
approaches~\cite{PopescuRohrlich,Vidal:2000,Horodecki} without suffering from
the disadvantages of other superadditive entanglement monotones. The one
proposed in~\cite{Eisert:Audenaert:Plenio}, for example, diverges on the set
of pure states.
\par
The one desirable property from the wish list of axiomatic entanglement
theory that we could not yet prove is continuity. We have shown, however,
that squashed entanglement is continuous near pure states and in the relative
interior of the faces of state space. Continuity in general would follow from
a conjectured Fannes type inequality for the conditional von Neumann entropy.
The proof of this conjecture thus remains the great challenge of the present
work. It might well be of wider applicability in quantum information theory
and certainly deserves further study.
\par
Another question to be asked is whether or not there exist states that are
nonseparable but, nonetheless, have zero squashed entanglement. We expect this
not to be the case: if not by means of proving that the infimum in squashed
entanglement is achieved, then by means of an approximate version of the result
of~\cite{SSA-eq}. The relation to entanglement measures other than entanglement
of formation, entanglement cost and distillable entanglment remains open in
general. If $\E=0$ would imply separability, however, it would follow that for
the class of PPT states, squashed entanglement is larger than entanglement
measures based on the partial transpose operation, like relative entropy of
entanglement, the logarithmic negativity and the Rains bound.

\acknowledgments
We thank C.~H.~Bennett for comments on an earlier version of
this paper. The work of MC was partially supported by a DAAD
Doktorandenstipendium. Both authors acknowledge support from the
U.K.~Engineering and Physical Sciences Research Council and the EU under
project RESQ (IST-2001-37559).

\appendix

\section{The classical case of the conditional Fannes inequality}
\label{app:classical:cond-Fannes}
In this appendix we prove the conjecture~\ref{cond-fannes} for states
\begin{align} \label{classical-states1}
  \rho^{AB}   &= \sum_k p_k \rho^A_k   \otimes \proj{k}^B, \\
 \label{classical-states2}  \sigma^{AB} &= \sum_k q_k \sigma^A_k \otimes \proj{k}^B,
\end{align}
with an orthogonal basis $( \ket{k} )_k$ and of ${\cal H}_B$,
probability distributions $(p_k)$ and $(q_k)$, and states
$\rho^A_k$ and $\sigma^A_k$ on $A$. Note that this includes the case
of a pair of classical random variables. In this case, the states $\rho^A_k$
and $\sigma^A_k$ are all diagonal in the same basis $( \ket{j} )_j$ of ${\cal
H}_A$ and thus $\rho^{AB}$ and $\sigma^{AB}$ describe joint probability
distributions on a cartesian product.
\par
The key to the proof is that for states of the form (\ref{classical-states1}),
$$S(A|B)_\rho= \sum_k p_k S\bigl(\rho^A_k\bigr),$$
and similarly for the states given in eq. (\ref{classical-states2}).
\par
First of all, the assumption implies that
$$\epsilon \geq \bigl\| \rho^B-\sigma^B \bigr\|_1 = \sum_k |p_k-q_k|.$$
Hence, we can successively estimate,
\begin{equation*}\begin{split}
  \bigl| S(A|B)_\rho-S(A|B)_\sigma \bigr|
                  &\leq \sum_k \left| p_k S\bigl(\rho^A_k\bigr)-q_k S\bigl(\sigma^A_k\bigr) \right| \\
                  &\leq \sum_k p_k \left| S\bigl(\rho^A_k\bigr)-S\bigl(\sigma^A_k\bigr) \right|     \\
                  &\phantom{=}
                       +\sum_k |p_k-q_k| S\bigl(\sigma^A_k\bigr)                                    \\
                  &\leq \sum_k p_k \bigl( \eta(\epsilon_k)+\epsilon_k\log d_A \bigr)                \\
                  &\phantom{=}
                       +\epsilon\log d_A                                                            \\
                  &\leq \eta(2\epsilon)+3\epsilon\log d_A,
\end{split}\end{equation*}
using the triangle inequality twice in the first and second lines, then using
$S(\sigma^A_k)\leq \log d_A$, applying the Fannes inequality,
lemma~\ref{fannes}, in the third (with $\epsilon_k:=\bigl\|
\rho^A_k-\sigma^A_k \bigr\|_1$), and finally making use of the concavity of
its upper bound.
To complete this step, we have to show $\sum_k p_k\epsilon_k \leq 2\epsilon$,
which is done as follows:
\begin{equation*}\begin{split}
  \epsilon &\geq \bigl\| \rho^{AB}-\sigma^{AB} \bigr\|_1
            =    \sum_k \bigl\| p_k\rho^A_k - q_k\sigma^A_k \bigr\|_1 \\
           &\geq \sum_k \biggl( \bigl\| p_k\rho^A_k - p_k\sigma^A_k \bigr\|_1
                              -\bigl\| p_k\sigma^A_k - q_k\sigma^A_k \bigr\|_1 \biggr) \\
           &\geq  \sum_k p_k\epsilon_k -\epsilon,
\end{split}\end{equation*}
where in the second line we have used the triangle inequality.
\qed
\par\medskip

Note that in the case of pure states the conjecture is directly implied by
Fannes inequality, lemma~\ref{fannes}, since $S(AB)=0$ and $S(A)=S(B)$.
Clearly, a proof of the general case cannot proceed along these lines as do
not have the possibility to present the conditional von Neumann entropy as an
average of entropies on $A$.

\end{document}